\documentclass[showpacs,aps,amsmath,amssymb,twocolumn]{revtex4}
\usepackage[latin1]{inputenc}
\usepackage{dcolumn}
\usepackage{bm}
\usepackage{verbatim} 
\usepackage{graphicx}			
\usepackage{color}
\newcommand{\be}{\begin{equation}}
\newcommand{\ee}{\end{equation}}
\newcommand{\ben}{\begin{eqnarray}}
\newcommand{\een}{\end{eqnarray}}
\newcommand{\bes}{\begin{subequations}}
\newcommand{\ees}{\end{subequations}}
\newcommand{\bb}{\bibitem}

\newcommand{\nn}{\nonumber\\}
\newcommand{\bfi}{\begin{figure}}
\newcommand{\efi}{\end{figure}}
\newcommand{\bc}{\begin{center}}
\newcommand{\ec}{\end{center}}
\newcommand{\sech}{\mbox{sech}}
\begin{document}
\title{First-order framework for flat brane with auxiliary fields}

\author{D. Bazeia$^{1,2}$, A. S. Lob\~ao Jr.$^{1}$, and R. Menezes$^{2,3}$}

\affiliation{$^1$Departamento de F\'\i sica, Universidade Federal da Para\'\i ba, 58051-970 Jo\~ao Pessoa, PB, Brazil\\ $^2$Departamento de F\'\i sica, Universidade Federal de Campina Grande, 58109-970 Campina Grande, PB, Brazil and \\ $^3$Departamento de Ci\^encias Exatas, Universidade Federal da Para\'\i ba, 58297-000 Rio Tinto, PB, Brazil.}

\begin{abstract}

This work deals with braneworld models in the presence of auxiliary fields. We investigate the case where
Einstein's equation is modified with the inclusion of extra, nondynamical terms. We show that the model supports first-order differential equations that solve the equations of motion, but the standard braneworld scenario changes under the presence of the parameter that controls the nondynamical or auxiliary fields that modifies Einstein's equation.

\end{abstract}

\pacs{11.27.+d, 11.10.Kk}

\maketitle

\section{introduction}

One important problem in the construction of alternative theories of gravity with the addition of extra dynamical fields is the presence of extra degrees of freedom which in general lead to instabilities in these theories \cite{ghost1,ghost2}. However, as shown very recently by Pani, Sotiriou, and Vernieri \cite{psv}, one may circumvent this problem introducing nondynamical or auxiliary fields. 

The modified gravity with nondynamical or auxiliary fields has then been studied in Refs.~\cite{c1,c2} within the cosmological context, to see how the auxiliary fields may contribute to the cosmic evolution. Moreover, it has also been recently studied within the thick braneworld context, in five dimensions with a single extra dimension of infinite extent \cite{gly}. There, the authors investigated the problem numerically, and found that the braneworld scenario in the presence of auxiliary field remains linearly stable.

The above investigations have motivated us to further study the thick braneworld scenario, but now extending the first-order framework developed in Ref.~\cite{A} to the presence of auxiliary fields, searching for exact solutions with a single extra dimension of infinite extent.

\section{The problem}

We start with a generalization of Einstein's equation, in the form
\begin{equation}\label{eq1}
R_{ab}-\frac12g_{ab} R=2T_{ab}+S_{ab}(T_{cd},g_{cd})\,,
\end{equation}
in five dimensional spacetime, where we are using $4\pi G^{(5)}=1$. Furthermore, the tensor $S_{ab}$ is obtained as a derivative expansion of the energy-momentum tensor $T_{ab}$ such that it vanishes when $T_{ab}=0$. We note that to keep the weak equivalence principle, we need $\nabla_a T^{ab}=0$; also, the Bianchi identity implies that $\nabla_a S^{ab}=0$.

Since the energy-momentum tensor $T_{ab}$ may contain second-order derivatives, up to fourth order in derivatives we can write \cite{psv}
\ben\label{eq2}
S_{ab}&=&\alpha_1 g_{ab}T+\alpha_2 g_{ab} T^2+\alpha_3 TT_{ab}+\nn
&&	+\alpha_4g_{ab}T_{cd}T^{cd}+\alpha_5 T^c{}_a T_{cb}+\nn
&&+\beta_1\nabla_a\nabla_b T+\beta_2 g_{ab}\Box T+\beta_3 \Box T_{ab}+\nn
&&+2\beta_4\nabla^c\nabla_{(a}T_{b)c}+\cdots
\een
where $T=g^{ab}T_{ab}$ and $\alpha_i$ and $\beta_j$ are real parameters. 

In this paper we search for exact solutions. This can be achieved if we consider only the lowest order term in the above equation \eqref{eq2}. Thus, we get the modified Einstein's equation
\ben\label{eq3}
R_{ab}-\frac12g_{ab} R&=&2\,T_{ab}+\alpha\, g_{ab}\,T\,,
\een
where we have changed $\alpha_1\to\alpha$.

The line element that describs a thick brane model in five-dimensional spacetime can be written as
\begin{equation}\label{eq4}
ds^2=g_{ab}dx^adx^b=e^{2A}\eta_{\mu\nu}dx^\mu dx^\nu -dy^2,
\end{equation}
where $\eta_{\mu\nu}$ is the metric of the four-dimensional flat spacetime, with signature $(+---)$. Of course, in the braneworld context with a single extra dimension of infinite extent, the warp function $A$ controls the warp factor $e^{2A}$ and is assumed to depend only on the extra dimension $y$, that is, $A=A(y)$.

We also assume that the background scalar field $\phi$ is defined by the standard Lagrange density, in the form
\begin{equation}
{\cal L}=\frac12 \nabla_a \phi \nabla^a \phi-V(\phi),
\end{equation}
where $V(\phi)$ is the potential that specifies the theory. Also, the scalar field only depends on the extra dimension, that is, $\phi=\phi(y)$. With this we can write the energy-momentum tensor in the form
\begin{equation}\label{eq5}
T_{ab}=g_{ab}\Big(\frac12 \phi^{\prime 2}+V\Big)+\nabla_a\phi\nabla_b\phi\,.
\end{equation}
In this case, the trace is $T=3\phi^{\prime 2}/2+5V$.

Using the line element \eqref{eq4} we can write the components of the Einstein equation \eqref{eq3} as
\be
6A^{\prime 2}= \frac{2-3\alpha}2 \phi^{\prime 2}-(2+5\alpha) V\,,\label{eq6b}
\ee
and
\be
-6A^{\prime 2}-3A^{\prime\prime}=\frac{2+3\alpha}2 \phi^{\prime 2}+(2+5\alpha) V\,,\label{eq6a}
\ee
or 
\be
A^{\prime\prime}= -\frac23\phi^{\prime 2}\,.\label{eq7}
\ee

To find exact solutions we follow the first-order formalism introduced in \cite{A}. For this we assume that the warp function $A(y)$ can be written in terms of another function, $W=W(\phi)$, in the form
\begin{equation}\label{eq8}
A^\prime=-\frac13 W\,,
\end{equation}
where $W$ is the superpotential \cite{DeWolfe:1999cp}. This is a first-order equation, which can be used to obtain, from \eqref{eq7}, another first-order equation 
\ben\label{eq9}
\phi^{\prime}=\frac12 W_\phi \,,
\een
where $W_\phi$ stands for $dW/d\phi$. We see from Eqs.~\eqref{eq8} and \eqref{eq9} that the two first-order equations are the same we get in the standard case, without auxiliary fields. However, in the present case the potential has to obey
\ben\label{eq10}
 V(\phi)=\frac{2-3\alpha}{8(2+5\alpha)} W_\phi ^2-\frac2{3(2+5\alpha)} W^2\,,
\een
and so it is changed by the modification introduced in Einstein's equation. One has to consider $\alpha>-2/5$, to avoid problem with the potential.  

Moreover, we can also write the energy density as
\begin{equation}\label{eq11}
\rho(y)=e^{2 A}\Big[\frac{2+\alpha}{4(2+5\alpha)} W_\phi ^2-\frac2{3(2+5\alpha)} W^2\Big]\,,
\end{equation}
and it is also changed by the $\alpha$ term. We note that both the potential and energy density reproduces the standard results, for $\alpha=0$; see, e.g., Ref.~\cite{Bazeia:2013euc}. 

We can start with a given $W(\phi)$ to obtain exact solutions for $\phi=\phi(y)$ and $A=A(y)$. Although the solutions are the same we get in the case of a standard thick brane with the same $W(\phi)$, we note here that the model is different, and the extra parameter $\alpha$ may be used to modify the physical effects in the braneworld scenario.

\section{Examples}

We now consider some specific examples. The first one is characterized by the polynomial function
\begin{equation}\label{eq12}
W(\phi)=2\phi-\frac23\phi^3\,,
\end{equation}
In the absence of gravity, this choice represents the wellknown $\phi^4$ model, with spontaneous symmetry breaking that has kink solution in the form
\begin{equation}\label{eq13}
\phi(y)=\tanh(y)\,.
\end{equation}
The warp function is obtained by integrating the equation \eqref{eq8}. It can be written as
\begin{equation}\label{eq15}
A(y)=\frac49 \ln(S)+\frac{S^2}{9}-\frac19\,,
\end{equation}
where $S={\rm sech}(y)$.

\begin{figure}[t]
\includegraphics[scale=0.6]{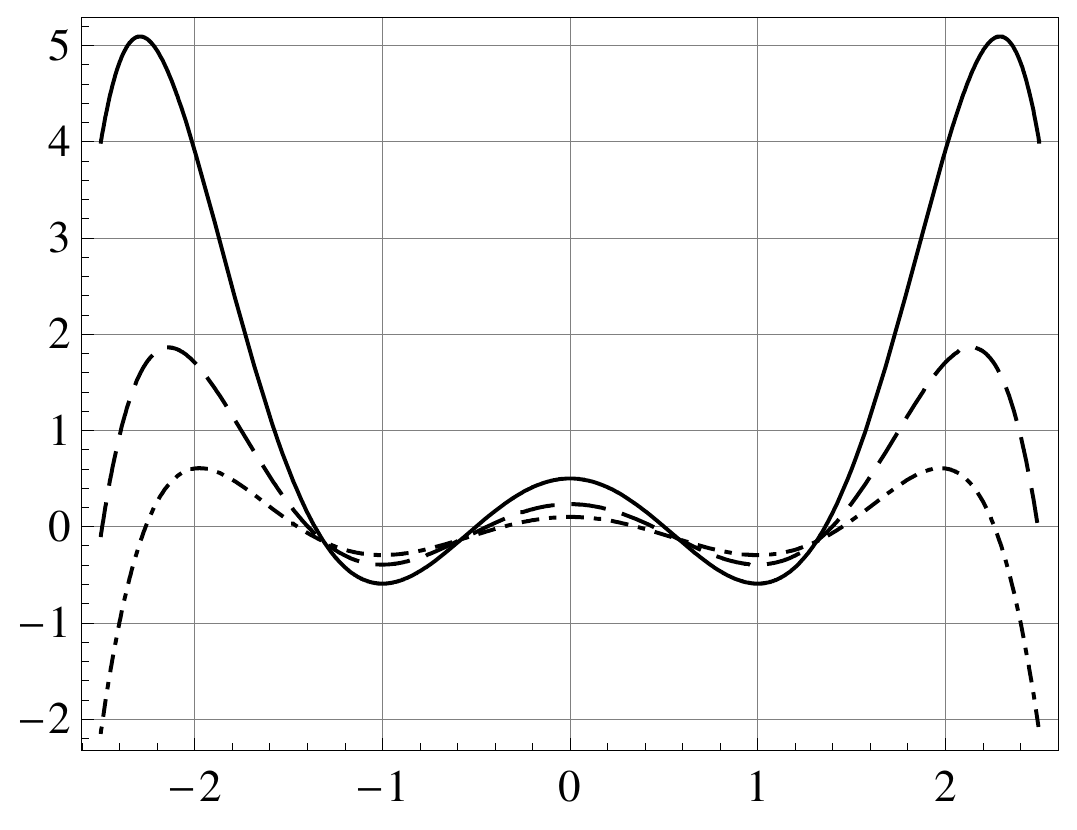} 
\caption{The potential \eqref{eq14}, depicted for $\alpha=0$ (solid line), $\alpha=0.2$ (dashed line) and $\alpha=0.4$ (dot-dashed line).}\label{fig1}
\end{figure}

\begin{figure}[t] 
\includegraphics[scale=0.62]{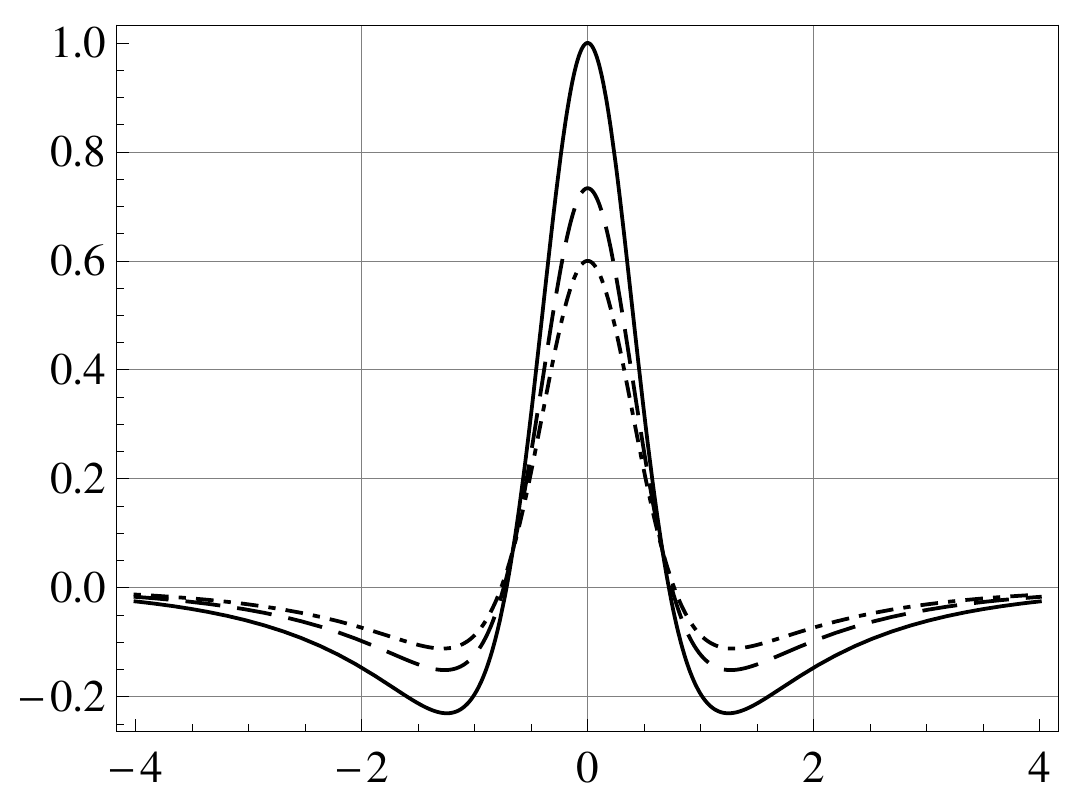} 
\caption{The energy density \eqref{eq16}, depicted for $\alpha=0$ (solid line), $\alpha=0.2$ (dashed line) and $\alpha=0.4$ (dot-dashed line).}\label{fig2}
\end{figure}

With the choice \eqref{eq12} we can write the potential \eqref{eq10} as
\ben\label{eq14}
 V(\phi)\!=\!\frac{2\!-\!3\alpha}{2(2\!+\!5\alpha)} \!\left(1\!-\!\phi^2\right)^2\!\!-\!\frac{8~\phi^2}{3(2\!+\!5\alpha)} \!\left(1\!-\!\frac13\phi^2\right)^2.
\een
The potential \eqref{eq14} is shown in Fig.~\eqref{fig1} for some values of the parameter $\alpha$. We see that the two local minima of the potential in $\phi\!=\!\pm 1$ give $V_{\min}\!=\!-32/(54\!+\!135\alpha)$; also, the maximum at $\phi\!=\!0$ is shifted as $V_{\max}\!=\!(2\!-\!3\alpha)/(4\!+\!10\alpha)$. Moreover, the energy density has the form
\begin{equation}\label{eq16}
\rho(y)=-\frac{e^{-2/9 + 2S^2/9}}{27 (2+5\alpha)}S^{8/9}\Big[32-3(26+9\alpha) S^4-8S^6\Big]\,.
\end{equation}
It is depicted in Fig.~\ref{fig2}, for some values of $\alpha$.

The second example concerns the sine-Gordon model, which is defined by the following function
\begin{equation}\label{eq17}
W(\phi)=2\sin\Big(\sqrt{\frac23}\phi\Big)\,.
\end{equation}
Here the kinklike solution has the form
\begin{equation}\label{eq18}
\phi(y)=\sqrt{\frac32}\arcsin\Big[\tanh\Big(\frac23 y\Big)\Big]\,.
\end{equation}

\begin{figure}[h] 
\includegraphics[scale=0.6]{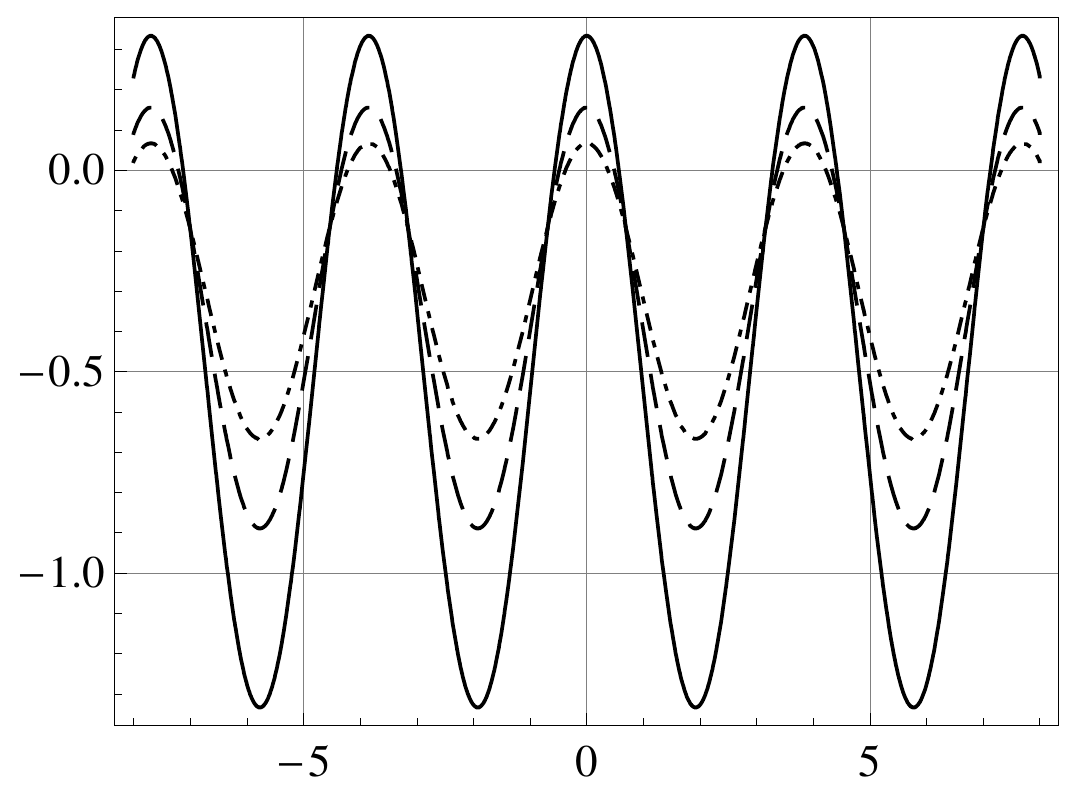} 
\caption{{The potential \eqref{eq20}, depicted for $\alpha=0$ (solid line), $\alpha=0.2$ (dashed line) and $\alpha=0.4$ (dot-dashed line).}}\label{fig3}
\end{figure}

\begin{figure}[h] 
\includegraphics[scale=0.65]{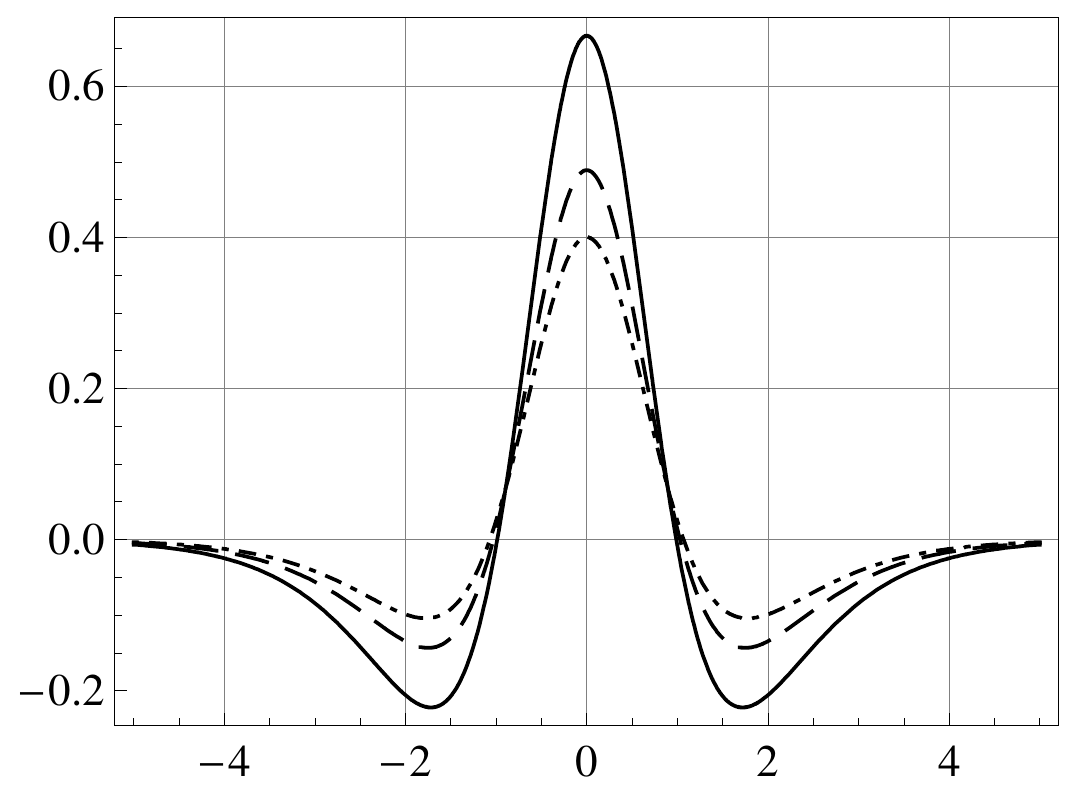} 
\caption{The energy density \eqref{eq21}, depicted for $\alpha=0$ (solid line), $\alpha=0.2$ (dashed line) and $\alpha=0.4$ (dot-dashed line).}\label{fig4}
\end{figure}

For the sine-Gordom model we get the warp function in the form
\begin{equation}\label{eq19}
A(y)=\ln\Big[\sech\Big(\frac23 y\Big)\Big]\,.
\end{equation}
Also, we can write the potential as
\ben\label{eq20}
V(\phi)=-\frac{8}{6+15\alpha}+\frac{10-3\alpha}{6+15\alpha} \cos^2\Big(\sqrt{\frac23}\phi\Big)\,.
\een
This potential has degenerate minima for each value of $\alpha$, as we illustrate in Fig.~\ref{fig3}. Moreover,
the corresponding energy density is
\begin{equation}\label{eq21}
\rho(y)=-\frac{2\, \sech^2\Big(\frac{2}{3}y\Big)}{6+15\alpha}\Big[4-(6+\alpha)\sech^2\Big(\frac{2}{3}y\Big)\Big]\,,
\end{equation}
which is depicted in Fig.~\ref{fig4}, for some values of $\alpha$.

\section{Stability} 

The investigation of linear stability of the gravity sector of the braneworld model can be done assuming that the metric is perturbed in the form
\be\label{eq22}
ds^2=e^{2A(y)}\Big[\eta_{\mu\nu}+h_{\mu\nu}(y,x)\Big] dx^\mu dx^\nu -dy^2\,.
\ee
Furthermore, the scalar field is written in the form
\be\label{eq23}
\phi=\phi(y)+\tilde \phi(y,x)\,.
\ee

The linear contributions to Einstein's equations are
\ben
\!\!\!&&\!\!\!\frac12 h_{\mu\nu}^{\prime\prime}\!+\!2A^\prime h_{\mu\nu}^\prime\!-\!\frac12 e^{-2A}\Box h_{\mu\nu}\!-\!\frac12e^{-2A}\eta_{\mu\nu}\partial^\alpha\partial^\beta h_{\alpha \beta}\!+\!\nn
\!\!\!&&\!\!\!+\frac12e^{-2A}\Big[\partial_\mu\partial^\alpha h_{\alpha \nu}+\partial_\nu\partial^\alpha h_{\alpha \mu}-\partial_\mu\partial_\nu h\Big]-\nn
\!\!\!&&\!\!\!-\eta_{\mu\nu}\Big[\frac12 h^{\prime\prime}\!+\!2A^\prime h^\prime\!-\!\frac12 e^{-2A}\Box h\!+\!\frac{2\!+\!5\alpha\!+\!12\alpha^2}{2\!+\!5\alpha}A^{\prime\prime}h\!+\!\nn
\!\!\!&&\!\!\!+\frac{8(1\!+\!\alpha)}{2\!+\!5\alpha}A^{\prime 2}h\Big]\!=\! \Big[(2\!+\!5\alpha) V_\phi \xi \!+\! (2\!+\!3\alpha) \phi^{\prime}\xi^{\prime}\Big]\eta_{\mu\nu},\label{eq24}
\een
where $h=h^{\mu}{}_\mu$.

In the gravity sector, we can simplify the investigation of stability considering the transverse traceless components of metric fluctuations, that is,
\begin{equation}\label{eq25}
\partial^\mu h_{\mu\nu}=0;\;\;\;\;\;h=0\,.
\end{equation}
Thus, we can check that Eq.~\eqref{eq24} reduces to the form
\ben\label{eq26}
\Big(\partial^2_y + 4 A^\prime \partial_y - e^{-2A}\Box\Big)h_{\mu\nu}=0\,.
\een

We introduce the $z$-coordinate in order to make the metric conformally flat, with $dz=e^{-A(y)}dy$ and we write
\ben\label{eq27}
H_{\mu\nu}(z)=e^{-ip\cdot x}e^{3/2 A(z)}h_{\mu\nu}\,.
\een
In this case, the 4-dimensional components of $h_{\mu\nu}$ obey the Klein-Gordon equation and the metric fluctuations of the brane solution lead to the Schroedinger-like equation
\ben\label{eq28}
\left[-\partial_z^2 + U(z) \right]H_{\mu\nu} = p^2 H_{\mu\nu}\,,
\een
where 
\ben\label{eq29}
U(z)=\frac{9}{4} A_z^2 + \frac32 A_{zz}\,.
\een
In this case we can write
\be
-\partial_z^2 + U(z) =S^\dagger S\,,
\ee
where
\be
S=\frac{d}{dz}-\frac32 A_z.
\ee
The operator $-\partial_z^2 + U(z)$ is then non-negative, and the gravity sector is linearly stable.
We note that the stability behavior in the gravity sector only depends on the warp factor $A(z)$. Since the warp factor does not depend on
$\alpha$, the modification introduced by the auxiliary field does not modify linear stability. For example, for the sine-Gordon model \eqref{eq17} we can write the quantum potential explicitly as
\begin{equation}\label{eq30}
U(z)=-6\frac{9-10z^2}{(9+4z^2)^2}\,,
\end{equation}
which was studied before in \cite{Gremm:1999pj}, in the absence of auxiliary fields.

\section{Ending comments}\label{end} 

In this work we extended the first-order framework developed in \cite{A} to the new scenario, where Einstein's equation is generalized to include non-dynamical or auxiliary fields. Our results show that the first-order equations for $\phi(y)$ and $A(y)$ are
\be
\phi^\prime=\frac12 W_\phi,\;\;\; \;\;A^\prime=-\frac13W,\nonumber
\ee
and have the very same form we get in the standard scenario. Although they remain unchanged, the potential has to obey
\ben
 V(\phi)=\frac{2-3\alpha}{8(2+5\alpha)} W_\phi ^2-\frac2{3(2+5\alpha)} W^2\,.\nonumber
\een
It explicitly depends on $\alpha$, the parameter that controls deviation from Einstein's equation. Thus, the energy density also depends on
$\alpha$. It is formally given by
\be
\rho=e^{2A}\left(\frac{2+\alpha}{4(2+5\alpha)} W_\phi ^2-\frac2{3(2+5\alpha)} W^2\right).\nonumber
\ee
We then see that although both $\phi(y)$ and $A(y)$ have the very same form they get in the absence of auxiliary fields, the potential and the energy density change in this new scenario. 

This is significant achievement, since the results suggest a new braneworld scenario with robust gravity sector which, differently from the standard braneworld scenario, engenders an extra parameter, $\alpha$, real. This parameter can be used to control physical aspects of the problem under study. Several issues can be investigated, and here we mention the splitting of the brane, an interesting effect that appears in specific models, as in the one-field model described in Ref.~\cite{BS1} and in the two-field or Bloch brane model described in Ref.~\cite{BS2}. We are now investigating how the extra parameter $\alpha$ of the current model may contribute to enhance the splitting of the brane. 

We also mention the hierarchy problem and the addition and entrapment of other fields to the brane \cite{h0,h1,h2,h3,h4}. Here we understand that the parameter $\alpha$ may be of use to drive the model towards more realistic possibilities. In addition, the proposed extension can be used to drive the model into new cosmological scenarios, with $\alpha$ playing interesting role; see, e.g., Ref.~\cite{c2}. In the cosmological context, baryogenesis is an important issue, and the current study motivates us to approach baryogenesis with the present extension, as also suggested in Ref.~\cite{cb}.

The authors would like to thank CAPES and CNPq for partial financial support.


\end{document}